# Data-Driven Transferred Energy Management Strategy for Hybrid Electric Vehicles via Deep Reinforcement Learning

Hao Chen[a, b], Gang Guo[a,*], Bangbei Tang[c], Guo Hu[b], Xiaolin Tang[a], Teng Liu[a]

a, Department of Automotive Engineering, Chongqing University, Chongqing 400044, China;

b, Department of Product Design, Chongqing Normal University, Chongqing 401331, China;

c, School of intelligent manufacturing engineering, Chongqing University of Arts and Sciences，Chongqing 402160, China;

*Corresponding author (Email address: cqguogang@163.com). Tel: +86(023)65362274; Fax: +86(023) 65362274.

**Abstract**: Real-time applications of energy management strategies (EMSs) in hybrid electric vehicles (HEVs) are the harshest requirements for researchers and engineers. Inspired by the excellent problem-solving capabilities of deep reinforcement learning (DRL), this paper proposes a real-time EMS via incorporating the DRL method and transfer learning (TL). The related EMSs are derived from and evaluated on the real-world collected driving cycle dataset from Transportation Secure Data Center (TSDC). The concrete DRL algorithm is proximal policy optimization (PPO) belonging to the policy gradient (PG) techniques. For specification, many source driving cycles are utilized for training the parameters of deep network based on PPO. The learned parameters are transformed into the target driving cycles under the TL framework. The EMSs related to the target driving cycles are estimated and compared in different training conditions. Simulation results indicate that the presented transfer DRL-based EMS could effectively reduce time consumption and guarantee control performance.

## 1. Introduction

Hybrid electric vehicles (HEVs) and plug-in hybrid electric vehicles (PHEVs) express great talent in energy-saving and emission-reducing [1]. Academic researchers are devoting themselves to improve the property of these vehicles, and car manufacturers are developing the new shapes of these vehicles. Energy management strategy (EMS) is a blue-chip study topic for these renewable vehicles. Owing to the ability to design reasonable distributions of multiple power resources, HEVs and PHEVs are able to promote energy efficiency and retard the greenhouse effect [2]. How to achieve an applicable online or rea-time EMS is still a challenging problem in the energy management field.

The approaches utilized to derive the EMSs are primarily classified into three types, which are rule-oriented, optimization-based, and learning-enabled methods [3]. Rule-oriented EMS is often built from the engineering experiences of human beings, and this policy is concentrated in real hybrid vehicles [4]. Optimization-based EMS is usually constructed from the optimization control techniques, and this strategy is always regarded as the benchmark to evaluate other policies [5]. Learning-enabled EMS is motivated by the intelligent regulation of machine learning, and this criterion is prevalent in recent years due to the online implemented potential [6]. Many methods have been conducted to found the learning-enabled EMSs, such as reinforcement learning (RL) and deep reinforcement learning (DRL).

RL is a conceptual structure to address the sequential decision-making problem through the intersection of an intelligent agent and its environment [7]. Since 2015, more and more attempts have been executed in the study of RL-based EMS for hybrid electric powertrains. For example, Liu et al. proposed a RL-based adaptive energy management policy for a hybrid tracked vehicle in [8]. The related performance is compared with the stochastic dynamic programming (SDP) method in different driving schedules. The authors in [9] formulated an advanced EMS for power-split PHEV with a heuristic planning reinforcement learning (RL) method. The Dyna-H algorithm is exploited to drive the control policy, which is proven to show advantages in fuel economy and computational speed. Furthermore, integration of the RL algorithm and other information is a novel pattern to generate real-time EMS. In Ref. [10], Zhou et al. summarized the combination of future driving conditions and RL to obtain the predictive EMS. The future driving information includes vehicle speed, acceleration, and driver behaviors. This information could enable the derived EMS to adapt to future driving environments and promote energy efficiency. The researchers in [11] discussed the energy management problem of HEV in a connected environment with proximal policy optimization (PPO) algorithm. It implies that the driving information (e.g., speed and acceleration) of the surrounding vehicles are known by the studied vehicle via V2V or V2I technology. Ref. [12] combined the onboard learning algorithm for Markov Chain, speedy Q-learning algorithm, and induced matrix norm to produce the online EMS. The relevant control policy is demonstrated to be superior to SDP and traditional RL in real-time driving situations.

DRL technique is the incorporation of deep learning and RL. It has achieved several impressive achievements in the energy management field in recent three years [13]. The main difference between RL and DRL is the implementation of deep learning to approximate the Q-table in RL. This feature is capable of promoting the accuracy and effectiveness of the conventional RL algorithms. For example, deep Q-learning (DQL) or deep Q-network (DQN) is the first adoptive approach in HEV’s energy management [14]. This algorithm could resolve the continuous optimal control problem with more precise powertrain modeling. In addition, some improved variations of common DQL are also applied in energy management research. The authors in [15] discussed the energy management problem with double DQL. This method is able to eliminate the overestimation issue in the approximation of Q-table effectively. Qi et al. [16] used dueling DQL to analyze the energy management problem. This technique employed state-value function and advantage function to acquire the Q-table, which could recognize the worth of each control action. Moreover, the deep deterministic policy gradient (DDPG) algorithm is another popular trial to dispose of the continuous state space and control space in energy management problems. Ref. [17] applied Gaussian process, random forest, and gradient boosted random trees to optimize the hyperparameters of the DDPG algorithm. Simulation results depict that the random forest could elevate the control performance of DDPG. To accelerate the convergence rate of random control actions selection, the authors in [18] embedded the optimal brake specific fuel consumption curve of the engine and the charge-discharge power characteristics of the battery in the training process of DDPG. This work shows that the particular powertrain efficiency map could speed up the learning efficiency and reduce fuel consumption.

Transfer learning (TL) is another learning paradigm of machine learning methods [19]. The concept of TL is transforming the learned knowledge from the source tasks to target tasks to accelerate the learning process. This method is inspired by the fact that humans could apply the learned knowledge previously to solve the new problems in high efficiency. TL is testified to be beneficial in many research fields. For example, the authors in [20] developed a traffic forecasting structure by combining the TL with a diffusion convolutional recurrent neural network. The provided predictive framework learned from the California highway network can forecast the traffic on the unseen regions of the network with high accuracy. To overcome the problem of insufficient training data in emotion recognition, Nguyen et al. [21] presented a TL-based strategy to learn cross-domain features to improve recognition performance. The availability and robustness of the proposed framework is verified. Ref. [22] discussed the algorithm recommendation problem with TL. The trained neural network is reused to transform the learned knowledge to similar datasets and the performance is guaranteed.

Recently, the integration of RL (or DRL) and TL is a promising solution to address the control problems in automotive engineering. Several applications have been carried out for the study of HEV or autonomous vehicles. For example, the authors in [23] combined DDPG and TL to construct a bi-level control architecture for a hybrid powertrain. The upper-level utilized DDPG to train the EMS at different speed intervals, and the lower-level employed TL to transform the pre-trained neural network for a new driving cycle. Lian et al. [24] realized cross-type knowledge transfer in the DRL-based EMS for hybrid electric vehicles. Four types of hybrid powertrains are included, and the EMS learned from Prius modeling is transformed into three other powertrains, which are a series, a series-parallel, and a power-split one. Furthermore, the knowledge transfer idea is also reflected in autonomous vehicles. Isele et al. discussed the decision-making problems at the intersection of self-driving vehicles [25]. The driving situations are cast into left-turning, going forward, and right-turning. The learned control policy from one driving task is proven to be effective in another driving mission. Ref. [26] exploited the transfer RL to determine the target recognition problem for the multi-autonomous underwater vehicles (AUV). The proposed transferred model could reduce the repeated calculation of similar data and ensure the real-time performance of the algorithm.

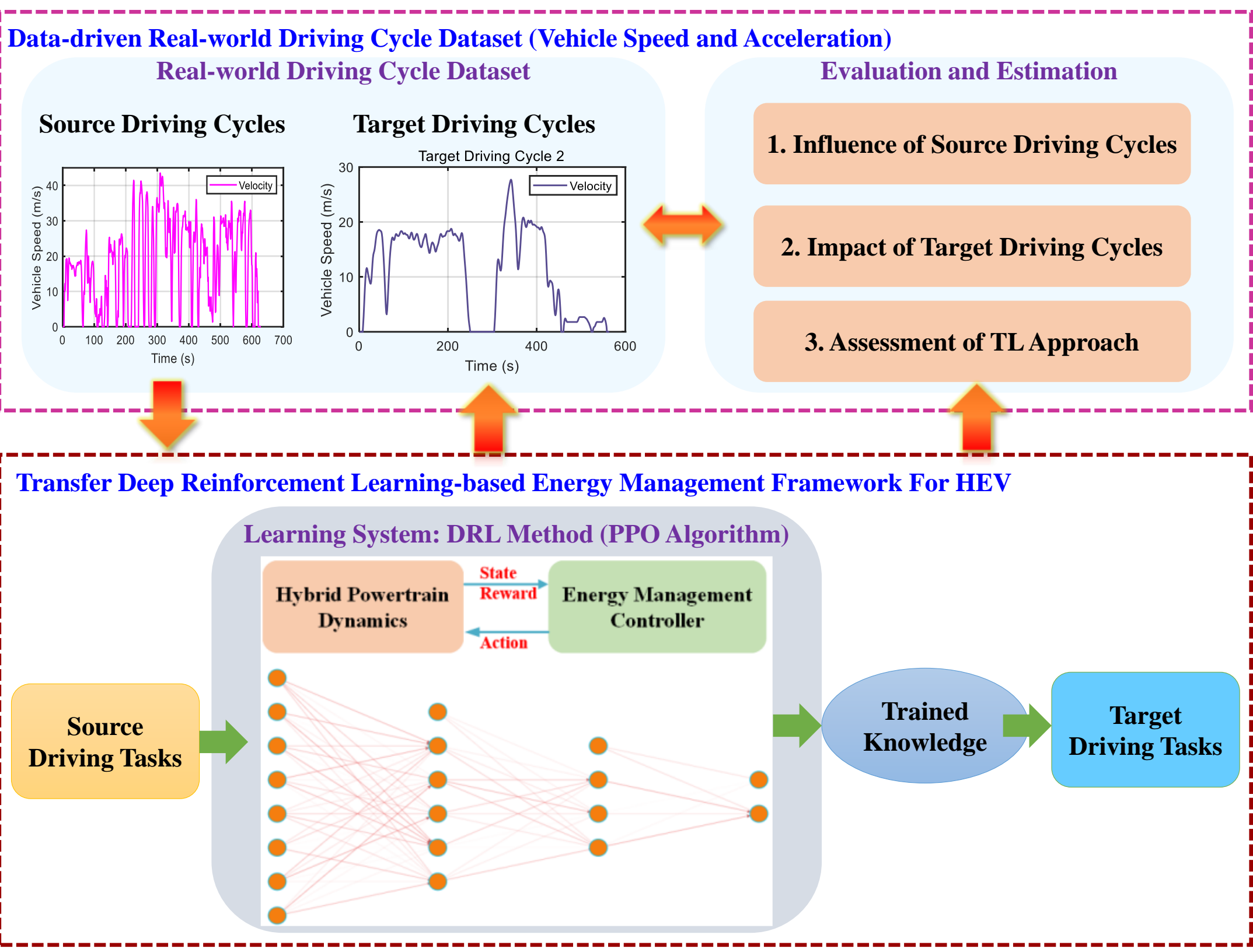

Fig. 1. The proposed transfer DRL-based control framework for data-driven energy management strategy.

Motivated by the merits of the TL and RL methods, this brief presented a data-driven transferred energy management strategy for hybrid electric vehicles, see Fig. 1 as an illustration. The special DRL algorithm is proximal policy optimization (PPO), which is located in the policy gradient approaches. The relevant EMS is built and analyzed in the real-world driving cycle dataset named Transportation Secure Data Center (TSDC). These driving cycles are classified into source driving cycles and target ones. The parameters in the deep learning are trained and learned from the source driving cycles, and then transformed into the target driving cycles to improve the learning efficiency. The impacts of the number of driving cycles and the significance of the testing driving cycle on the related EMS are illustrated. Multiple assessments are executed to evaluate the efficacy of the transfer DRL-based energy management policy. Simulation results indicate that the presented transfer DRL-based EMS could effectively reduce time consumption and guarantee control performance.

Three perspectives of the main contributions and innovations are given in this work. 1) a real-time energy management policy is constructed for a HEV by using DRL (PPO algorithm) and TL; 2) Real-world driving cycle dataset is applied to formulate and estimate the data-driven EMS; 3) the necessity and importance of the source and target driving cycles are illuminated to promote the control performance of the proposed method. This paper is an attempt to combine the TL, DRL, and real-world driving data to found the real-time EMS for a HEV. This presented control framework is adaptive to different hybrid powertrains.

The rest content of this work is organized as follows. The powertrain modeling of the studied HEV and its relevant energy management problem is described in Section 2. Section 3 gives the implemented progress of the transfer DRL approach, including the PPO and TL algorithms. In Section 4, the simulation results of different testing experiments are elaborated. The effects of the source driving cycles are discussed. Finally, Section 5 summarizes the key takeaways.

## 2. Hybrid Powertrain Modeling and Energy Management Problem

Taking a series-parallel hybrid powertrain as an example, this section establishes the powertrain modeling and its related energy management problem. The sketch of this studied powertrain is depicted in Fig. 2 [27]. The main components include a battery pack, an internal-combustion engine (ICE), an integrated starter generator (ISG), and a traction motor. The power distribution between the battery and ICE is treated as the optimization objective in order to improve energy efficiency. Since the core of this work focuses on the application of transfer DRL in energy management problems, battery aging, road slope, and temperature effect are not considered in this article. The presented EMS is easily generalized to the different hybrid powertrain.

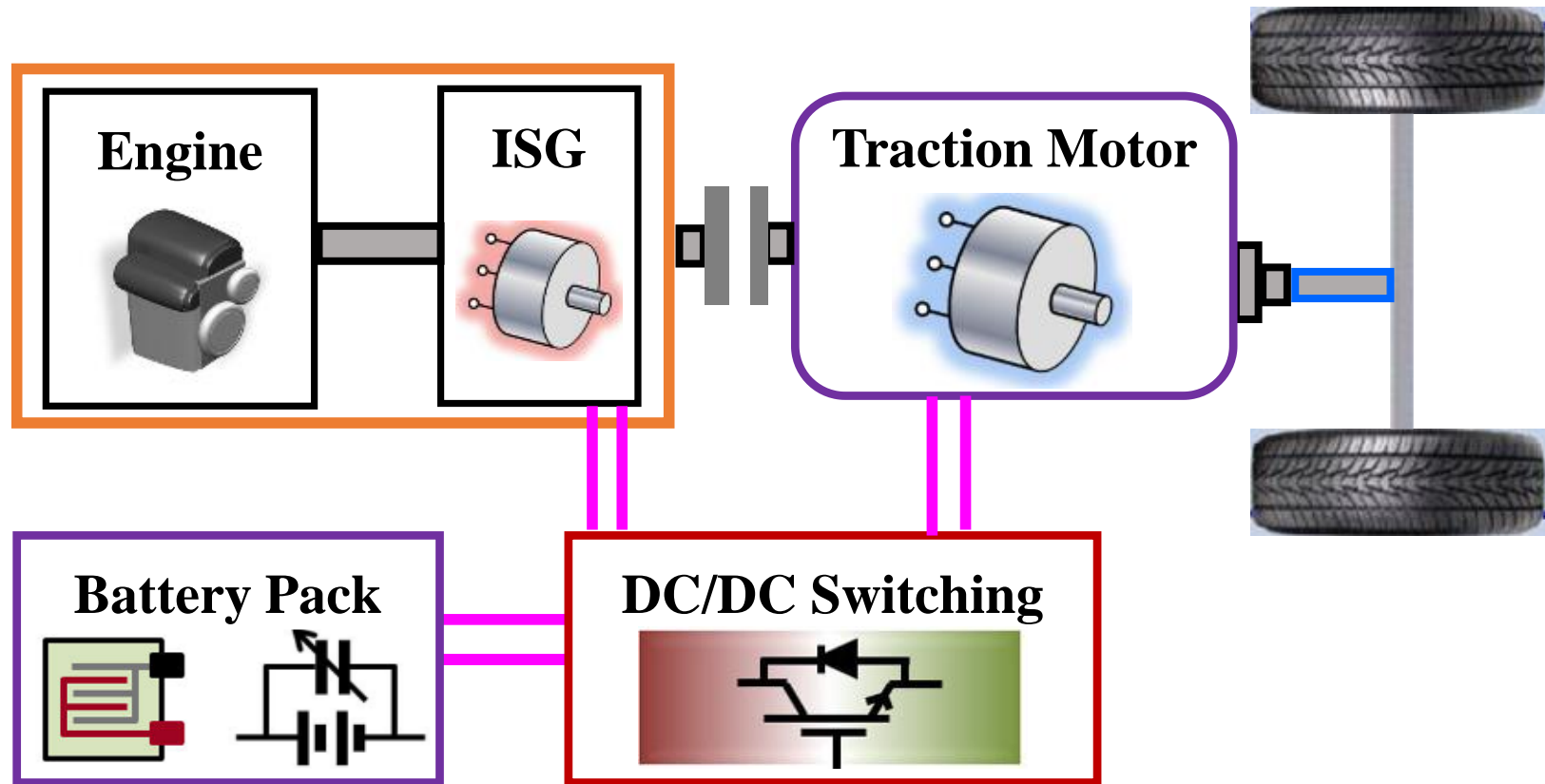

Fig. 2. Control architecture of the research series-parallel powertrain topology [27].

### 2.1 Modeling of the Hybrid Powertrain

The power request of the wheels is a significant parameter in the energy management problem, because it decides the amount of power provided by the ICE and battery. In the backward modeling of

HEV, this power request is constituted by three parts as follows:

$$P_{req} = P_{ro} + P_{ae} + P_{ine} \tag{1}$$

where $P_{ro}$, $P_{ae,}$ and $P_{ine}$ are the allied powers according to the rolling resistance, aerodynamic drag, and inertial force, respectively. They can be expressed further in the following format:

$$\begin{cases} P_{ro} = m_v \cdot g \cdot f \cdot v \\ P_{ae} = \frac{1}{2} \cdot \rho \cdot A_a \cdot C_D \cdot v^2 \cdot v \\ P_{ine} = m_v \cdot a \cdot v \end{cases} \tag{2}$$

where $m_v$ is the vehicle mass, $f$ is the rolling resistance coefficient, $g$ is the gravity coefficient. $\rho$ is the air density, $A_{a,}$ and $C_D$ are the frontal area and aerodynamic drag coefficient, respectively. $v$ and $a$ are the vehicle speed and acceleration. They are varied with the driving cycles and thus lead to the mutable power request. In this work, the driving cycles are adopted from the real-world driving cycle dataset, which would be introduced in Section 2.3.

In general, the battery and ICE are the main energy sources to maintain the power request of the powertrain. Hence, the dynamic modeling of these two components is necessary for the energy management problem. In the battery, state-of-charge (SOC) is a principal argument to determine the remaining electricity capacity at each time instant. The value of SOC ranges from 0 to 1 to represent the fully charged and fully discharged conditions of the battery. It could be computed as follows:

$$\dot{SOC} = -I_{bat} / Q_{cap} \tag{3}$$

where $I_{bat}$ is the battery current, and $Q_{cap}$ is the nominal capacity of the battery. In this work, the battery is mimicked by the internal resistance model [28], wherein the power and voltage of the battery are represented by the battery current as:

$$\begin{cases} P_{bat} = U_{bat} \cdot I_{bat} \\ U_{bat} = V_{oc} - I_{bat} r_0 \end{cases} \tag{4}$$

where $V_{oc}$ is the open-circuit voltage in the battery, and $r_0$ is the internal resistance. For the particular series-parallel hybrid powertrain, the nominal capacity $Q_{cap}$ is 8.1Ah, the nominal voltage of 200V, and the internal resistance $r_0$ is 0.25Ω. Incorporating (3) and (4), the differential of SOC is depicted as:

$$\dot{SOC} = -(V_{oc} - \sqrt{V_{oc}^2 - 4r_0 P_{bat}}) / (2Q_{cap} r_0) \tag{5}$$

SOC could reflect the current conditions of the battery, and thus it is chosen as the state variable of the energy management problem. The range of SOC is [0, 1] in this article because the real-world driving cycles are considered. As the power request is satisfied by the battery and ICE, the power of ICE is selected as the control action and it is capable of deciding the battery power at each time step.

In the static map modeling of ICE, the significant factor is the fuel consumption rate. It is computed by the speed and torque of ICE as follows:

$$\dot{m}_{ICE} = f(T_{ICE}, \omega_{ICE}) \tag{6}$$

where $\dot{m}_{ICE}$ is the instantaneous fuel consumption, $T_{ICE}$ and $\omega_{ICE}$ are the torque and speed of ICE, respectively. $f$ in (6) often indicates the look-up table method, which is supported by the brake specific fuel consumption (BSFC) curve of ICE.

The parameters of the ICE are labeled as follows: the range of the ICE speed is [1000, 4500] rpm, the peak power is 57 kW at 5000 rpm, and peak torque is 115 Nm at 4200 rpm [27]. The control action in the energy management problem is the ICE torque with continuous space, which is expected to be resolved by the advanced DRL method. The parameters of this research series-parallel hybrid powertrain are described in Table 1.

Table 1. Factors of the series-parallel powertrain for energy management problem.

| Indicator | Meaning | Values |
|---|---|---|
| $m_v$ | Vehicle mass | 1325 kg |
| $\rho$ | Air density | 1.225 kg/m$^3$ |
| $f$ | Rolling resistance coefficient | 0.012 |
| $A_a$ | Frontal area | 2.16 m$^2$ |
| $C_D$ | Aerodynamic drag coefficient | 0.26 |
| $g$ | Gravity coefficient | 9.8 m/s$^2$ |
| $V_{oc}$ | Open circuit voltage | 150 V |
| $SOC_{ref}$ | Charge sustaining value | 0.65 |

## 2.2 Energy Management Problem

In this work, the energy management problem of HEV is transformed into an optimization control problem with specific optimization objective and control constraints. The control goal is the sum of two parts. The former is the instantaneous fuel consumption in (6) to represent the fuel economy target. The latter is the SOC restraint to balance the final value of SOC after running an exclusive driving cycle. This goal is named cost function and written as follows:

$$\begin{cases} J = \int_0^T [\dot{m}_{ICE}(t) + \lambda(\Delta_{SOC}(t))^2]dt \\ \Delta_{SOC}(t) = \begin{cases} SOC(t) - SOC_{ref} & SOC(t) < SOC_{ref} \\ 0 & SOC(t) \geq SOC_{ref} \end{cases} \end{cases} \tag{7}$$

where [0, $T$] is the time horizon of each driving cycle. $\lambda$ is a positive weighting factor in achieving a trade-off between these two control indexes, and it is settled as 1000 in this paper. $SoC_{ref}$ is a pre-defined parameter to realize the charge-sustaining module, and it equals to 0.65 in this article.

Since the original intention of the energy management problem is searching the optimal control actions at each time instant, the state variables and control actions should be defined. The state variables are vehicle speed, acceleration, and SOC in the battery. The control action is the ICE torque with continuous space. They are limited as follows:

$$\begin{cases} S=\{v \in [0, 45]m/s, a \in [-5, 5]m/s^2, SOC \in [0, 1]\} \\ A=\{T_{ICE} \in [0, 115]Nm\} \end{cases} \tag{8}$$

In the selected process of control policy, many variables should work in a reasonable physical range. It indicates that the parameters in the ICE, battery, ISG, and traction motor should follow a couple of constraints. These restraints are decided by the different hybrid powertrain. In the studied series-parallel powertrain, these conditions are stated as follows:

$$\begin{cases} SOC_{\min} \leq SOC(t) \leq SOC_{\max} \\ P_{bat,\min} \leq P_{bat}(t) \leq P_{bat,\max} \end{cases} \tag{9}$$

$$\begin{cases} \omega_{x,\min} \leq \omega_x(t) \leq \omega_{x,\max}, & x = Mot, Gen, ICE \\ T_{x,\min} \leq T_x(t) \leq T_{x,\max}, & x = Mot, Gen, ICE \end{cases} \tag{10}$$

where min and max imply the minimum and maximum value of each variable. The subscript of *Mot*, *Gen,* and *ICE* represent the speed or torque in each component. The next subsection gives the source of the real-world driving cycle dataset.

## 2.3 Driving Cycle Dataset

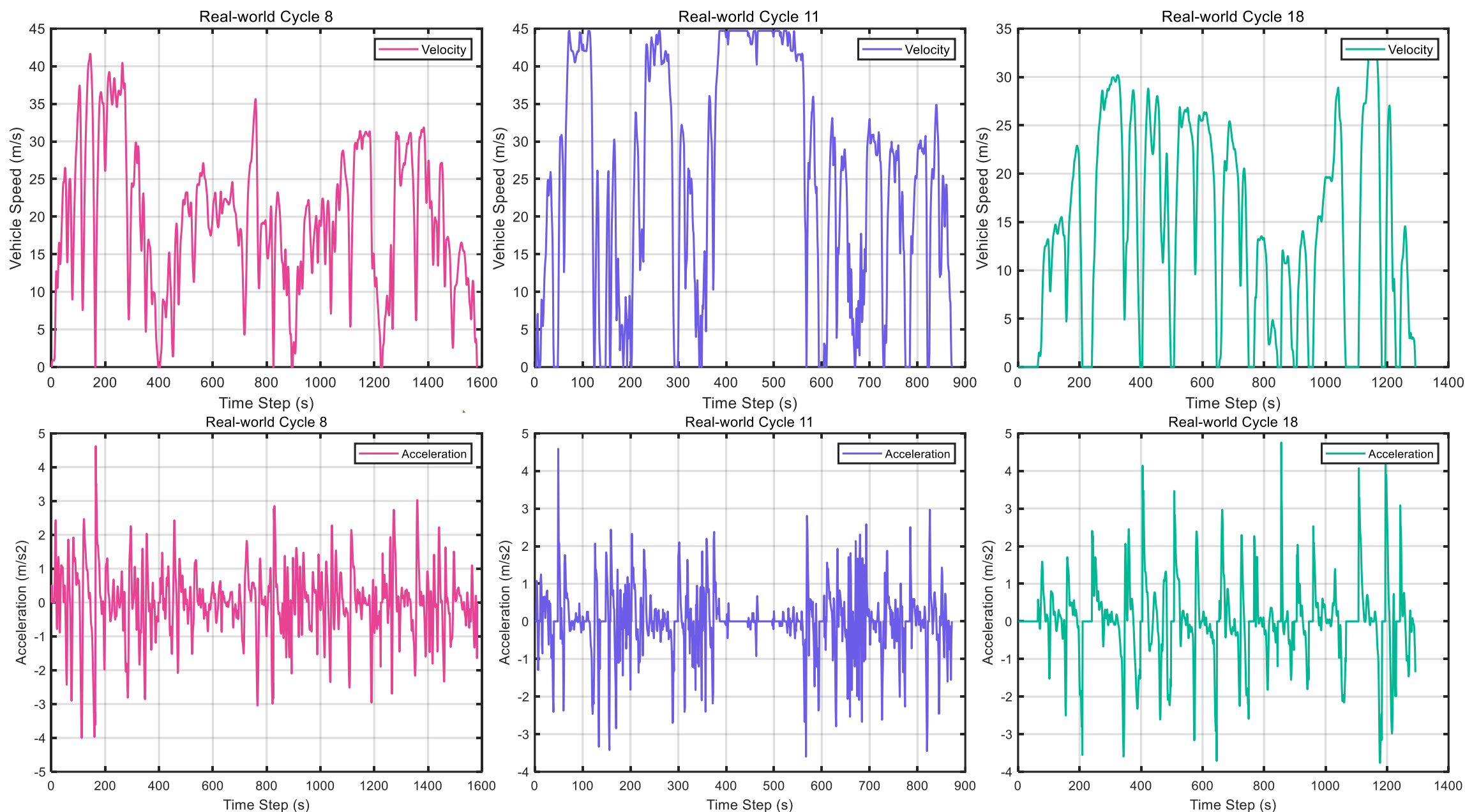

Fig. 3. Selected real-world driving cycles for transfer DRL-based EMS research.

In most references related to the EMS of HEV, the simulation driving cycles are the standard cycles [29]. As this work aims to propose a real-time EMS for multiple hybrid powertrains, the real-world driving cycles are adopted. The concrete driving cycle dataset is collected by Transportation Secure Data Center (TSDC) [30]. This driving cycle data includes aggregated statistics, trip driving distances, and second-by-second profiles (speed and acceleration) collected from vehicles with global positioning systems (GPS). The involved types of automobiles are light-, medium-, and heavy-duty vehicles. The data processing procedures are introduced in [31]. The generated cleansed data are available for many open-source softwares for programming, database query, and analysis, such as PostgreSQL/PostGIS/QGIS, GRASS, Python, and R.

As explained in (2), vehicle speed and acceleration are the necessary elements for the power request calculation. The driving cycles in the energy management problem basically imply the various vehicle speed and acceleration profiles. In this work, the Delaware valley regional planning commission (DVRPC) processed driving cycles in TSDC are used to derive the real-time EMSs. For specification, 20 driving cycles (these cycles are labeled as Cycle 1 to Cycle 20) are comprised and analyzed by the proposed transfer DRL method. Fig. 3 shows the sampled vehicle speed and acceleration profiles. It can be discerned that the time duration, maximum value of speed, and average velocity are different in these driving cycles. As a consequence, the variation of state variables (e.g., SOC) is different in these control cases.

As mentioned in the Introduction, there are the source and target task domains in the TL method. In this paper, the transferred ability of the DRL-based EMS is investigated and analyzed. A part of the 20 driving cycles is categorized as the source driving cycles, and the DRL approach is forced on them to train the parameters of the neural network. Then, the trained neural networks are working on the target driving cycles to improve the learning efficiency and control performance. The influences of the ingredient of the source driving cycles would be discussed, and the knowledge of the target driving cycles will also be illuminated. In the next section, the particular DRL method (PPO algorithm) and the TL method are introduced.

## 3. DRL Method and TL Technique for EMS Transformation

This section describes the implementation process of the transfer DRL method. This approach is utilized to formulate the real-time EMS in this article. The core of the RL framework is given firstly, including the essential elements of one RL algorithm. Then, the proximal policy optimization (PPO) algorithm is introduced, which is suitable for the optimal control problem with continuous space. It is a novel form of policy gradient technique. Finally, the TL method is employed to transform the learned EMS from the source driving cycles to target driving ones.

### 3.1 RL Implementation

The RL framework narrates the interacted procedure of an intelligent agent and its environment. At each step, the agent takes a control action and the environment gives a state and feedback. According to the feedback from the environment, the agent could modulate and improve its control policy. In this work, the agent is the onboard energy management controller and the environment is the powertrain dynamics. The goal of the agent is minimizing the cost function in (7) for a given driving cycle by selecting an appropriate control strategy.

The intersection between the agent and environment is often mimicked as a Markov decision process (MDP), as a tuple ($\mathcal{S}$, $\mathcal{A}$, $\mathcal{P}$, $\mathcal{R}$). MDP means a decision-making process has Markov property, which indicates that the next state variable only depends on the present state, not on the past states. In the tuple, $\mathcal{S}$ is a set of state variables, and $\mathcal{A}$ is a set of control actions. $\mathcal{R}$ is a reward function, it delegates the feedback from the environment and is represented by the instantaneous cost function in (7). $\mathcal{P}$ is a transition model, which describes the development trace of the state variables.

RL method is exploited to formulate the EMS (a sequence of control actions) in the HEV's energy management problem. The chosen regulation of the control actions is to minimize the cumulative rewards. This accumulated reward is computed as the sum of the current reward and discounted future rewards as follows [32-34]:

$$R_t = \sum_{k=t}^{T} \gamma^{k-t} \cdot r_k \tag{11}$$

where $t$ is the time instant, $\gamma \in [0, 1]$ is a discount factor in balancing the current and future received rewards. This is because the control action in the RL framework would affect not only the present reward but also the future feedbacks. $r_k \in \mathcal{R}$ is the instantaneous reward.

Assuming the EMS (control strategy) obtained from the energy management controller is $\pi$, it is calculated based on the value functions via RL algorithms. Two value functions are usually applied to derive the control policy in RL algorithms. They are named as state-value function and action-value function, and they are written as the following format, respectively:

$$V_\pi(s_t) \doteq E_\pi\left[\sum_{k=t}^{T} \gamma^{k-t} \cdot r_k \mid s_t\right] \tag{12}$$

$$Q_\pi(s_t, a_t) \doteq E_\pi\left[\sum_{k=t}^{T} \gamma^{k-t} \cdot r_k \mid s_t, a_t\right] \tag{13}$$

where $s_t$ and $a_t$ indicating the state variables and control action at time step $t$. $\mathrm{E}_\pi$ [·] denotes the expected value of the accumulated reward following policy $\pi$. Different RL algorithms aim to use disparate criteria to update the value functions. As the control action is displayed in action-value function, many RL algorithms are inspired to update this function (called Q-matrix for short).

For convenience, the action-value function is rewritten as the recursive form to insulate the current reward as follows:

$$Q_\pi(s_t, a_t) = E_\pi\left[r_t + \gamma \min_{a_{t+1}} Q_\pi(s_{t+1}, a_{t+1})\right] \tag{14}$$

where $s_{t+1}$ and $a_{t+1}$ are the next state variable and control action. To compute the optimal control policy, the Bellman optimality equation is described with the transition model:

$$Q^{*}(s_t,a_t)=\sum_{s_{t+1},r_t} p(s_{t+1},r_t \,|\, s_t,a_t)\left[r_t+\gamma \min_{a_{t+1}} Q^{*}(s_{t+1},a_{t+1})\right] \tag{15}$$

where $p(s_{t+1}, r_t|s_t, a_t) \in \mathcal{P}$ means the transition probability from the current state to the next state. The tuple $(s_{t+1}, r_t, s_t, a_t)$ is called as an observation or a transition in the RL algorithm to record the stepwise intersection between the intelligent agent and its environment. Finally, the optimal control action at each step is generated from the trained Q-matrix as follows:

$$a_t^{*} = \arg\min_{a_t} Q(s_t,a_t) = \arg Q^{*}(s_t,a_t) \tag{16}$$

In the conventional RL algorithm, the Q-matrix is updated by filling the observation from each step. It is time-consuming and inefficient. Hence, in the DRL approach, the deep learning (e.g., neural network) is leveraged to approximate the Q-matrix. The policy is modeled with a parameter $\theta$, $\pi_\theta(a|s)$. The factor $\theta$ implies the parameters of the neural networks. In the next subsection, the PPO algorithm is introduced to update the parameterized policy.

3.2 PPO Algorithm

For a parameterized cost function $J(\theta)$ in (6), the gradient $\nabla_\theta J(\theta)$ could help the agent to find the best $\theta$ for $\pi_\theta$ that produces the lowest cumulative reward. This gradient can be computed with respect to the action-value function as follows [35]:

$$\begin{aligned}\nabla_\theta J(\theta) &= \sum_{s} p^{\pi}(s) \sum_{a} \nabla_\theta \pi_\theta(a \mid s) Q^{\pi}(s,a) \\ &= \sum_{s\in S} p^{\pi}(s) \sum_{a\in A} \pi_\theta(a \mid s) Q^{\pi}(s,a) \frac{\nabla_\theta \pi_\theta(a \mid s)}{\pi_\theta(a \mid s)} \\ &= E_{\pi}\left[Q^{\pi}(s,a) \nabla_\theta \ln \pi_\theta(a \mid s)\right]\end{aligned} \tag{17}$$

where $p^{\pi}(s)$ is the probability distribution of the state variable. $\pi_\theta(a \mid s)$ and $\pi_{\theta_{old}}(a \mid s)$ are the new and old policies, the probability ratio between two policies are depicted as:

$$r(\theta)=\frac{\pi_\theta(a \mid s)}{\pi_{\theta_{old}}(a \mid s)} \tag{18}$$

Then, the loss function of updating the policy in the RL framework is represented by:

$$L^{PG}(\theta) = \mathrm{E}\left[\ln \pi_\theta(a \mid s) \hat{A}(s,a)\right] \tag{19}$$

where $\hat{A}(s, a)$ is the estimated advantage function because the real one is usually unknown. In the conventional policy gradient method, the loss function $L^{PG}$ is often used to perform multiple steps of optimization with the same policy. However, this updating criterion may cause some problems, such as sample inefficiency, policy diversity, and hesitation in exploration and exploitation. The new parameter $\theta$ may make the policy alter too much at one step. Thus, the PPO algorithm is presented in [36] to overcome these challenges by synthesizing the advantages of typical value-based and policy-based RL methods.

To improve the stability of policy updating in PPO algorithm, a constraint is forced on the probability ratio $r(\theta)$ to stay within a small horizon around 1, which is [1-$\varepsilon$, 1+$\varepsilon$]. The new loss function is described as follows:

$$L^{CLIP}(\theta) = \mathrm{E}\left[\min(r(\theta)A_{\theta_{old}}(s,a), \mathrm{clip}(r(\theta),1-\varepsilon,1+\varepsilon))\hat{A}_{\theta_{old}}(s,a)\right] \tag{20}$$

where $\varepsilon$ is a small positive constant close to zero (equal to 0.2 in this work), which is a hyperparameter. The function clip($r(\theta)$,1− $\varepsilon$,1+$\varepsilon$) clips the probability ratio to be no more than 1+$\varepsilon$ and no less than 1−$\varepsilon$. This function could effectively avoid achieving a large policy updating from the old policy.

In the actual application of PPO, the neural network of the actor (policy) would share the parameters with that of the critic (state-value function). To encourage sufficient exploration for the agent, the clipped loss function is augmented with an error term on the value estimation and an entropy term:

$$L^{CLIP+VF+S}(\theta) = \mathrm{E}_{\pi}\left[L^{CLIP}(\theta) - c_1(V_\theta(s) - V_{target})^2 - c_2 H(s,\pi_\theta)\right] \tag{21}$$

where the second term is the error term, and the third term is the entropy term. $c_1$ and $c_2$ are the tuned coefficients, which are the hyperparameter. When updating the policy in the PPO algorithm, a $K$ timesteps ($K$ is less than the length of the episode) sample data is collected via recurrent neural networks. These data are used to calculate the loss function in (21), wherein the estimated advantage function is written as:

$$\begin{cases} \hat{A}_t(s,a) = \delta_t + (\gamma\lambda)\delta_{t+1} + \ldots + (\gamma\lambda)^{K-t+1}\delta_{K-1} \\ \delta_t = r_t + \gamma V(s_{t+1}) - V(s_t) \end{cases} \tag{22}$$

where $\lambda$ is another discount factor for the advantage function. In each iteration, there would be $M$ actors to collect the data. The implemented pseudo-code of the PPO algorithm is depicted in Table 2, which is proved to produce awesome results in some control missions [37-38]. In the energy management problem, the state variables and control action are specialized in (8). The transition model of vehicle speed and acceleration is provided by the real-world driving data. The transition model of SOC is represented by (5). The reward is the instantaneous cost function in (7).

Table 2. Implemented pseudo-code of the PPO algorithm.

```
PPO Algorithm, Actor-Critic Style
1. For iteration = 1, 2, …, do
2.     For actor = 1, 2, …, M do
3.         Run policy π_θold in environment for K timesteps
4.         Calculate advantage function based on (22), Â_1,…,Â_K
5.     end for
6.     Optimize loss function in (21) with respect to θ for Z epochs
7.     Update θ_old with θ
8. end for
```

The default parameters for the applied PPO algorithm are defined as follows: the discount factor $\gamma$ and learning rate $\alpha$ in the RL framework are 0.9 and 0.01. The timesteps $K$ is 512, the mini-batch size $Z$ is 64, the hyperparameter $\varepsilon$ is 0.2, and the discount coefficient for advantage function $\lambda$ is 0.92. The hyperparameter $c_1$ and $c_2$ in (21) are 0.5 and 0.01. After training, the PPO-based EMS is generated and the related parameters of the neural network are stored. For a new driving cycle, the learned parameters are transformed to accelerate the learning efficiency. The TL method is introduced in the next subsection.

### 3.3 TL Technique

The transfer EMS for HEV means transforming the learned neural networks from the source driving cycles to target driving cycles in this paper. The essence is reusing the parameters of the actor-critic network. For different driving cycles, the relevant EMSs should be disparate. However, the definition of reward, state variables, control action, and transition model is the same. The leading EMS is still the power split controls between the ICE and battery. Hence, the parameters of the actor-critic network could be transformed to improve learning efficiency.

The neural network related to the source driving cycles is named as an expert network, and the neural network related to the target driving cycles is the student network. In this work, the number of source driving cycles is larger than that of target driving cycles. The source driving cycles could comprise the target driving cycles or not. The weight and bias of the expert network would be first trained and obtained. These parameters are the initialized factors for training the EMS according to the target driving cycles. All the parameters of the actor and critic networks would be replaced. The hyperparameter stays the same in these two training processes. This operation is capable of maintaining the control performance and elevate the learning efficiency in the target driving cycles.

In this work, the number of the source driving cycles is defined as 5, 10, and 20, respectively. The effect of the training data in the source domain on the control performance will be evaluated. These driving cycles are all extracted from one dataset to guarantee the correlation. For a fixed number, the source driving cycles would be designed to contain the target driving cycles or not. This layout would estimate the knowledge of target driving cycles on the efficacy of EMS. In the next section, the proposed transfer DRL-based EMS is assessed in multiple simulation tests.

## 4. Simulation Results and Analyzation

This section evaluates the transfer PPO-enabled EMS for HEV in three aspects. First, the number of the source driving cycles are various, and the performance of the relevant EMS is discussed. Then, the existent of the target driving cycles is analyzed. It implies that the source driving cycles would include the target driving cycles or not. Finally, the PPO algorithm with and without the TL method are compared. This simulation experiment would estimate the TL for accelerating the learning efficiency.

### 4.1 Influence of Source Driving Cycles

This section discusses the impact of the number of source driving cycles on EMS of HEV. The number of the source driving cycles is defined as 5, 10, and 20, respectively. For convenience, the source driving cycles and target driving cycles in the transfer DRL-based EMS are shortened as SDC and TDC. Two real-world driving cycles are chosen as the TDC, and these TDC in the analysis of Section 4 are the same. Fig. 4 depicts the speed trajectories of the two TDC. The learned neural networks from different SDC are transformed to train the EMSs for these two TDC. In this subsection, the three cases of SDC include the TDC.

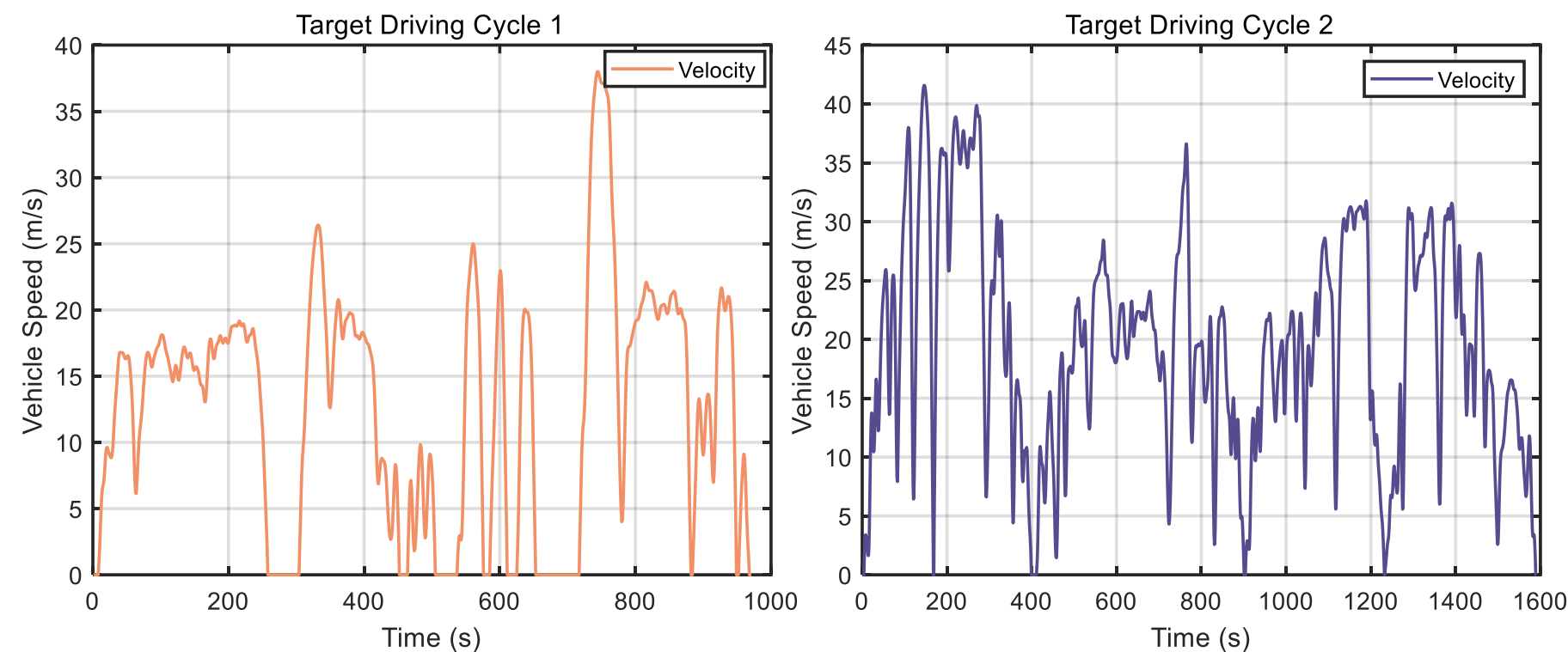

Fig. 4. Two target driving cycles used in the discussion and analysis in Section 4.

To describe the effectiveness of the constructed EMS using the proposed method, the state variables, and control action would be compared and analyzed. Since the vehicle speed and acceleration are extracted from the real-world driving data, Fig. 5 shows the SOC curves for two selected TDC. In each TDC, three control cases are enumerated. The main difference is the number of source driving cycles, which represents the source knowledge in the presented transfer DRL approach. From Fig. 5, it is evident that the SOC variations are not the same for these two TDC. For target driving cycle 1, for example, the SOC of 20 SDC is totally different from those in 5 SDC and 10 SDC in the time interval [400, 700]. For target driving cycle 2, the special time horizon is [400 800]. Moreover, it can be discerned that the vehicle speeds are very large around 750 in TDC 1 and around 200 in TDC 2. It means the power demand of the powertrain is big, and thus the ICE and battery should provide the power simultaneously. As a consequence, the SOC would decline at these time steps. It implies that the control logic of different EMSs is the same, but the particular control actions are different for different numbers of SDC.

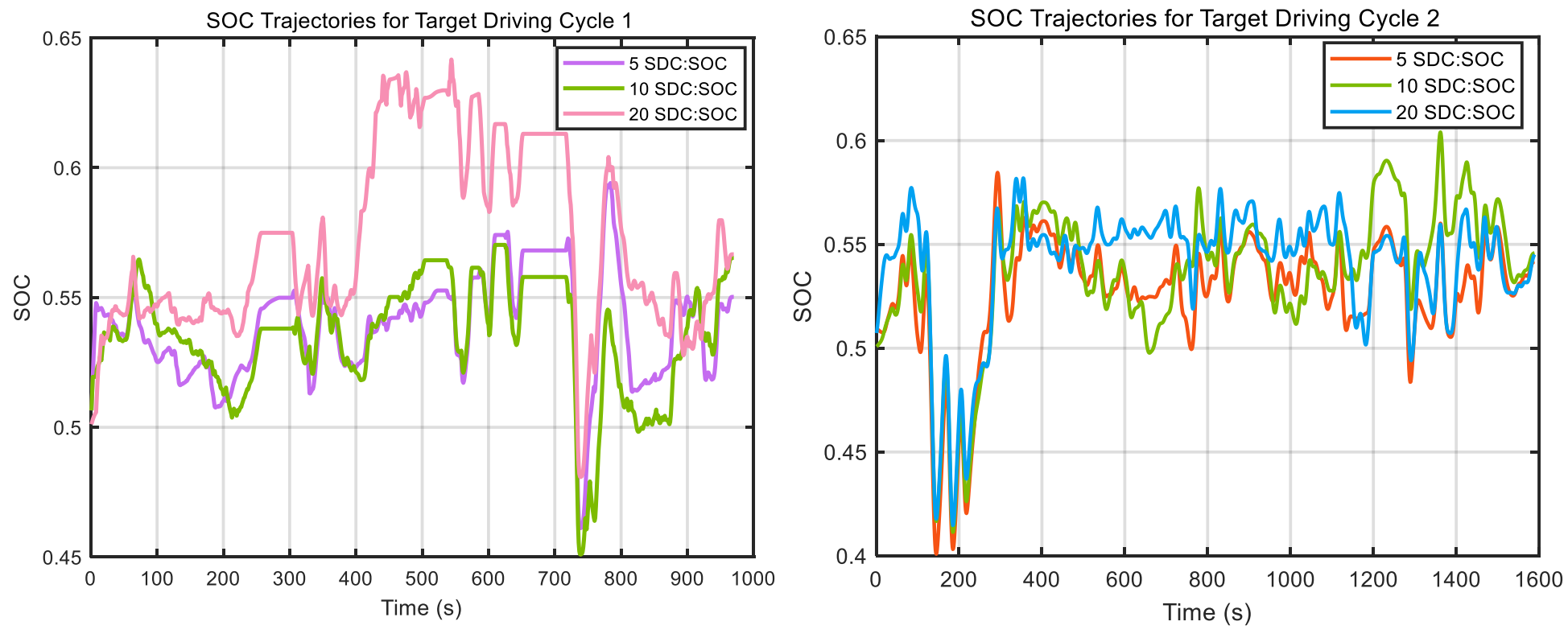

Fig. 5. SOC trajectories for two TDC in different control cases.

After exhibiting the SOC graphs, the power split controls between the ICE and battery are given in Fig. 6. The first line is for TDC 1, and the second line is for TDC 2. The powers of ICE in TDC 1 are usually lower than those in TDC 2. This is affected by vehicle speed in Fig. 4, in which the TDC 2 is often bigger than TDC 1. For TDC 1, the power splits with respect to different numbers of SDC are different. Especially in the time range [300 600] (shown in the purple rectangle), the ICE and battery powers are different. For TDC 2, the power distribution between ICE and battery is also not the same. As described in the purple ellipse, the variations of ICE are different in the time horizon [400 800]. This distinction is decided by the stated control action, ICE torque. Since the power demand is the same for TDC 1 or TDC 2, the ICE torque would influence the ICE power and thus affect the battery power. Furthermore, the SOC is determined by battery power, and thus the SOC curves are different in Fig. 5. This difference would also influence the reward, which will be shown in the next content.

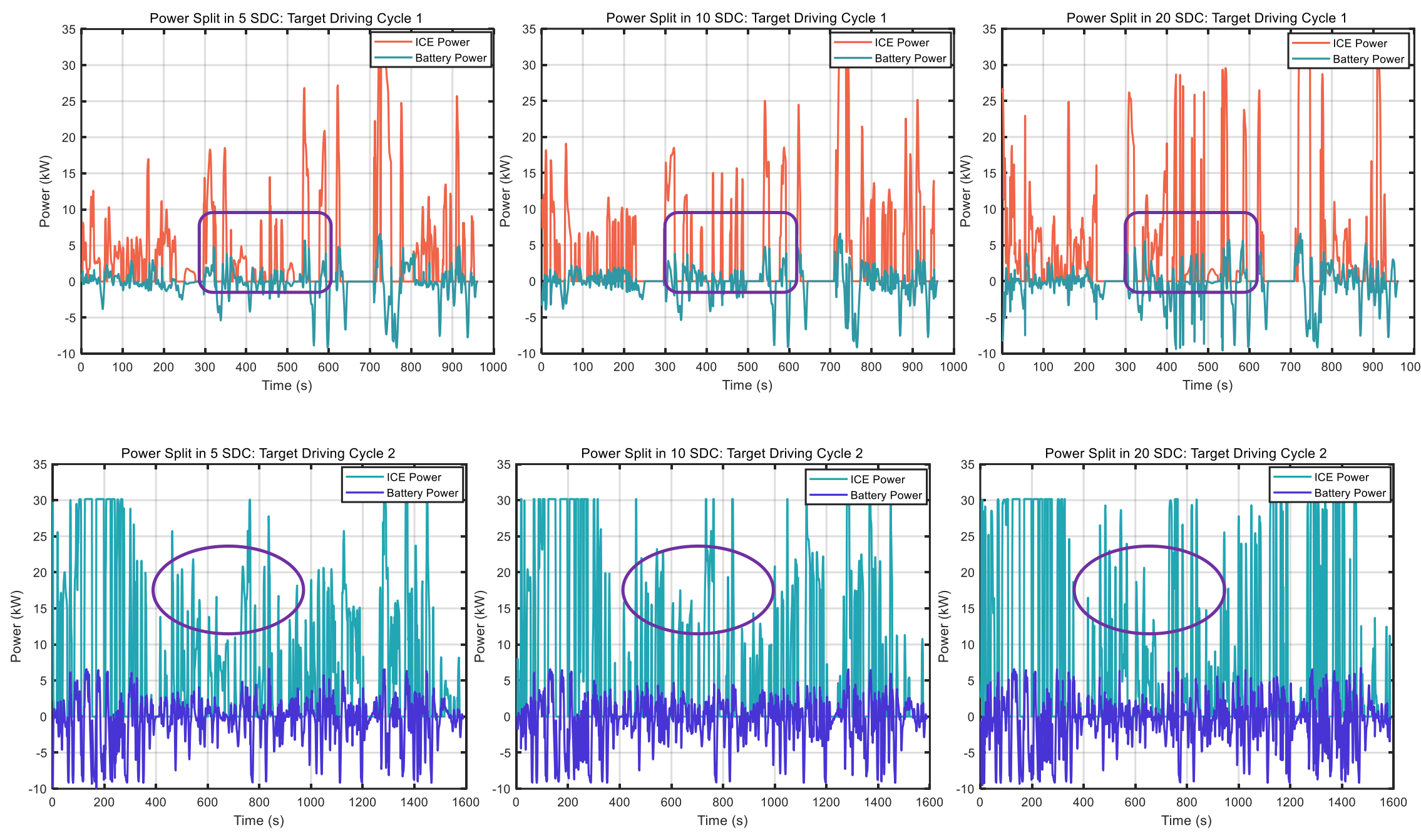

Fig. 6. Power distribution between ICE and battery in three control cases for two TDC.

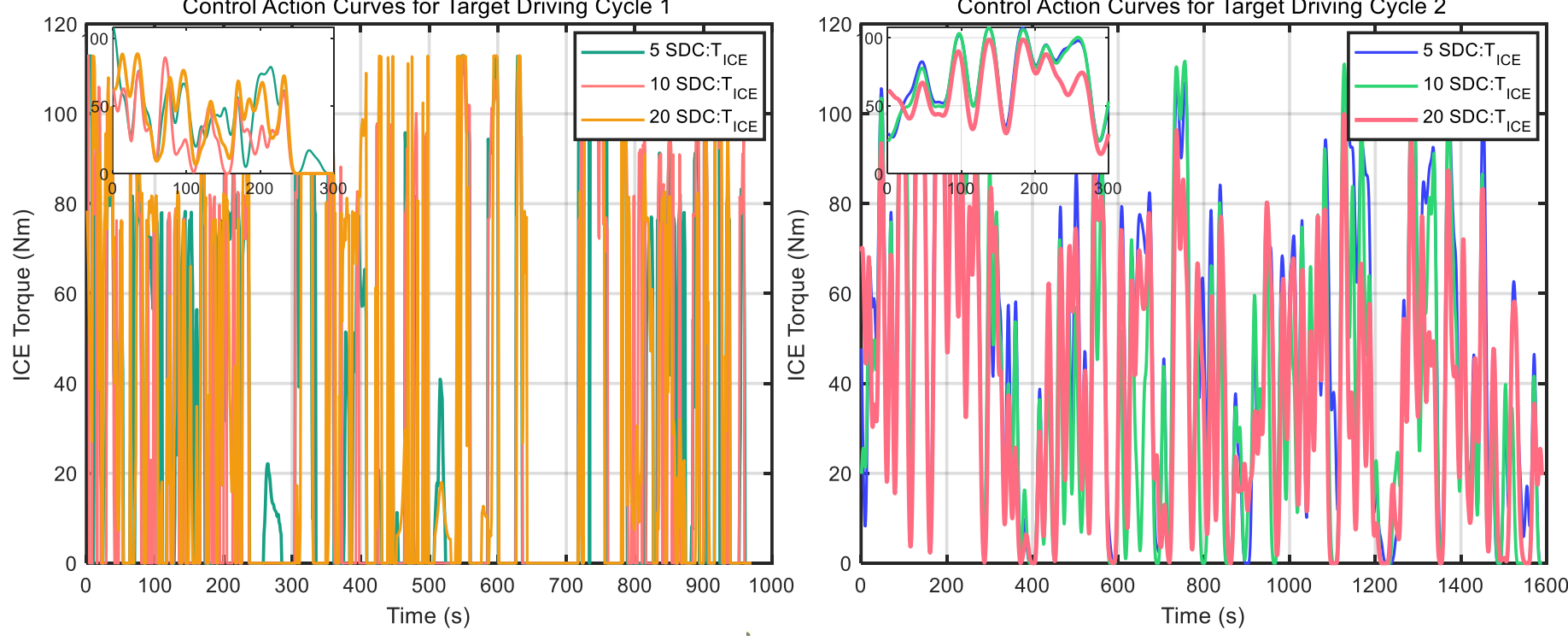

Fig. 7. Different control policies in the energy management problem for TDC 1 and TDC 2.

To intuitively illuminate the differences in control actions, Fig. 7 demonstrates the torque of ICE ($T_{ICE}$) among disparate control cases. For each TDC, the three control cases are labeled as 5 SDC, 10 SDC, and 20 SDC. It can be found that the control actions at some special time steps are different. When the ICE torque is zero, it implies that the ICE does not work. Thus, the working conditions of ICE are nearly the same for different cases of SDC. However, the output torques are diverse at the same time instant. This is attributed to the transformed neural networks from the SDC. Since the transformed knowledge is different, the obtained EMSs are not the same, which is reflected by the trajectories of control actions. The zoom-in figure could show the differences in the particular time interval [0 300] in these two TDC.

Finally, the total reward (shown in (7)) is the most concerning factor in the DRL-based energy management problem, Fig. 8 lists the total reward of different control cases in two TDC. In this graph, the X-axis indicates the different SDC situations, and the Y-axis represents the two TDC. Hence, the first row means the different control cases for TDC 1, and the second row is for TDC 2. In each TDC, the 20 SDC case leads to the highest total reward, which expresses the best control performance. Along with the increase in the number of SDC, the total reward would be higher. This feature is caused by the driving information contained by the SDC. More source driving cycles would result in a more mature neural network. Based on this neural network, the transfer EMS would be better. In conclusion, in the proposed transfer DRL-enabled EMS, a larger number of SDC is welcome. Thus, in the following discussion, the number of SDC is defined as 20, and this number stays the same in different control cases.

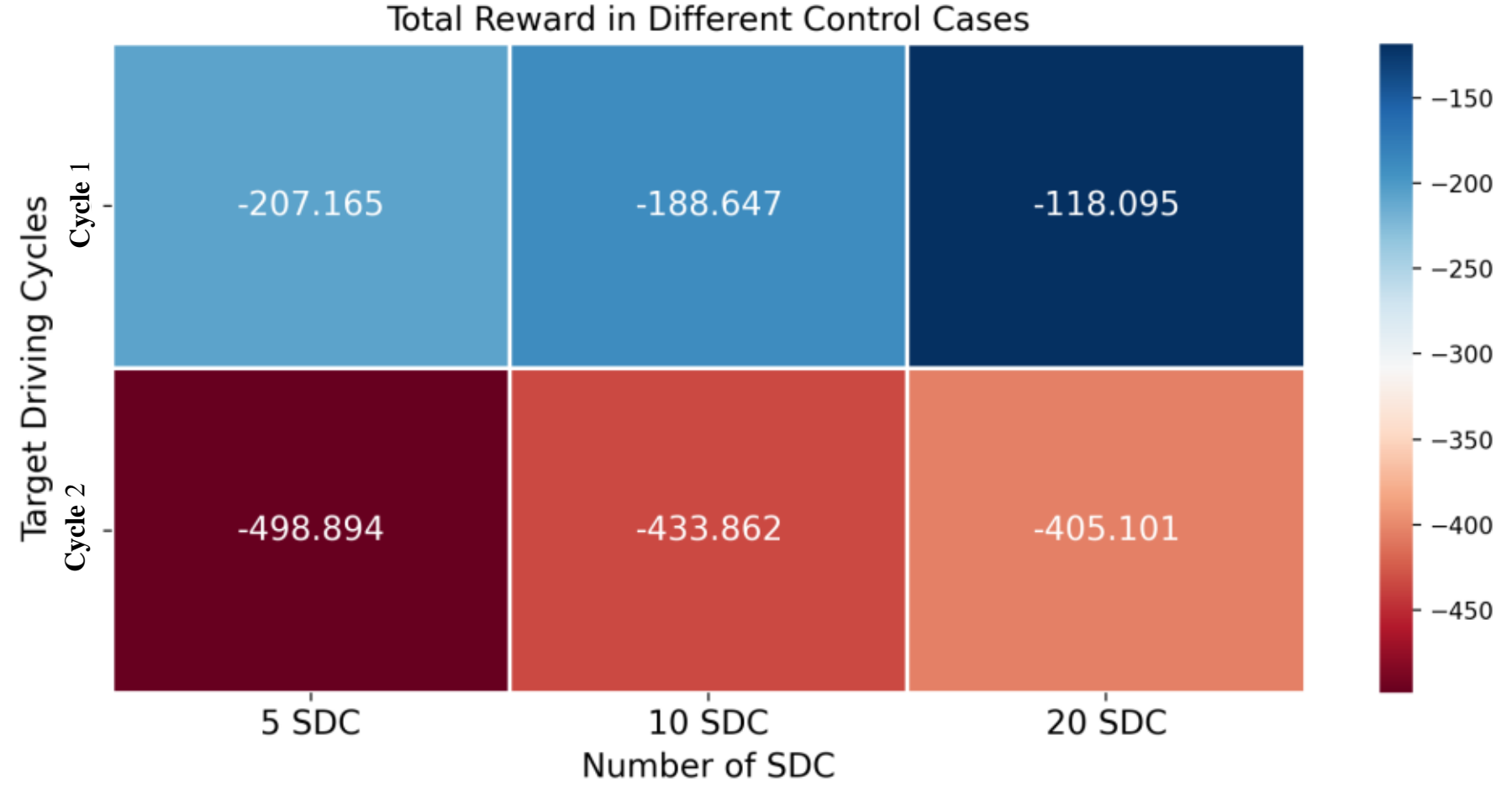

Fig. 8. Total reward with respect to different control cases for two TDC.

4.2 Impact of Target Driving Cycles

This section illuminates the influence of the existence of TDC in SDC. For specification, two control cases are compared, which are 20 SDC include the two TDC, or 20 SDC do not include TDC. If the SDC contain TDC, the learned and transformed neural networks are assumed to be more adaptive to the fixed TDC. The relevant simulation results would criticize this suggestion. The testing driving cycles are still the two TDC in Fig. 4.

First, the trajectories of the state variable (SOC) are displayed in Fig. 9. The variations of these graphs are not precisely the same. At time step 750 in TDC 1 and 200 in TDC 2, the SOC decreases sharply due to the enormous power request at the wheels. However, the SOC curves are different at many time instants. It is caused by the battery power directly, which declares that the control strategies in these two cases for each TDC are different. As a consequence, the TDC information in the SDC would affect the transformed neural networks. In actual driving, the information of target driving cycles is usually unknown to the energy management controller. Forecasting technology is capable of being considered to acquire future driving information.

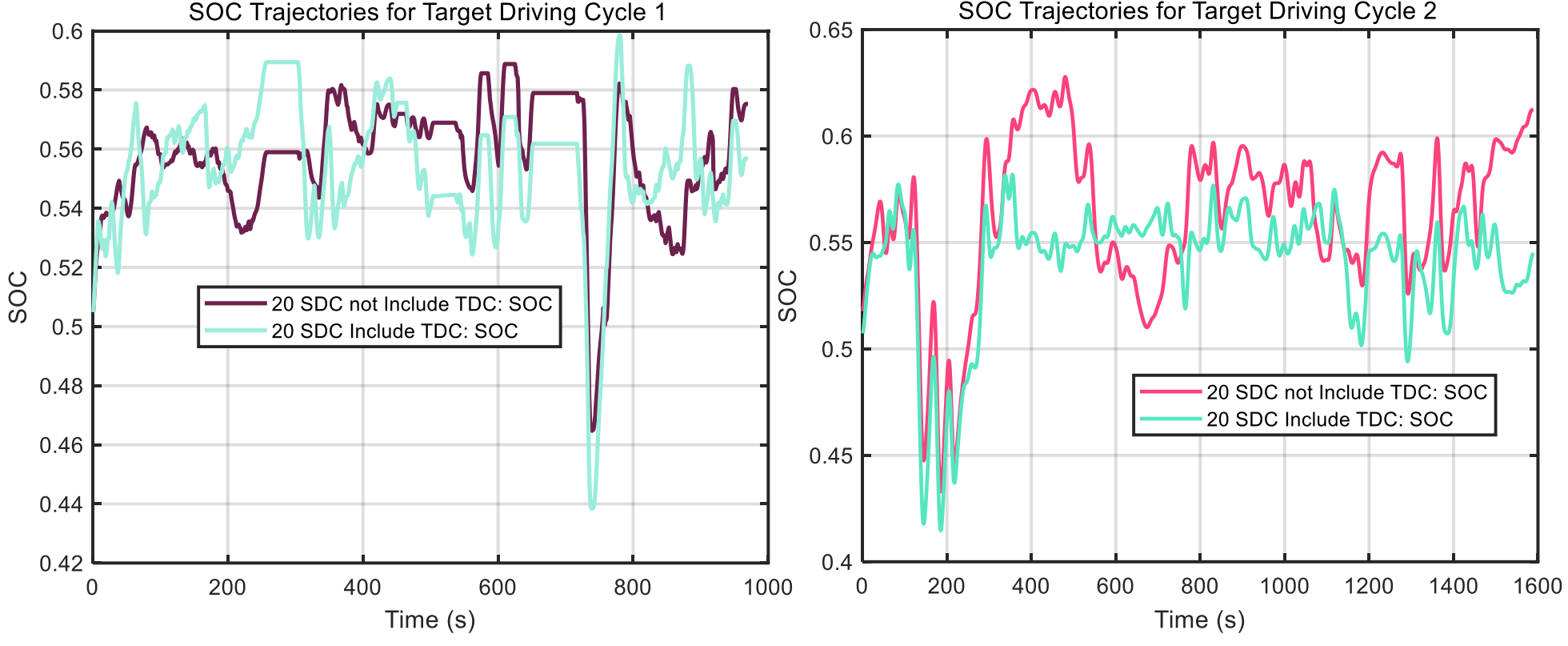

Fig. 9. SOC curves for evaluation of the significance of the TDC information.

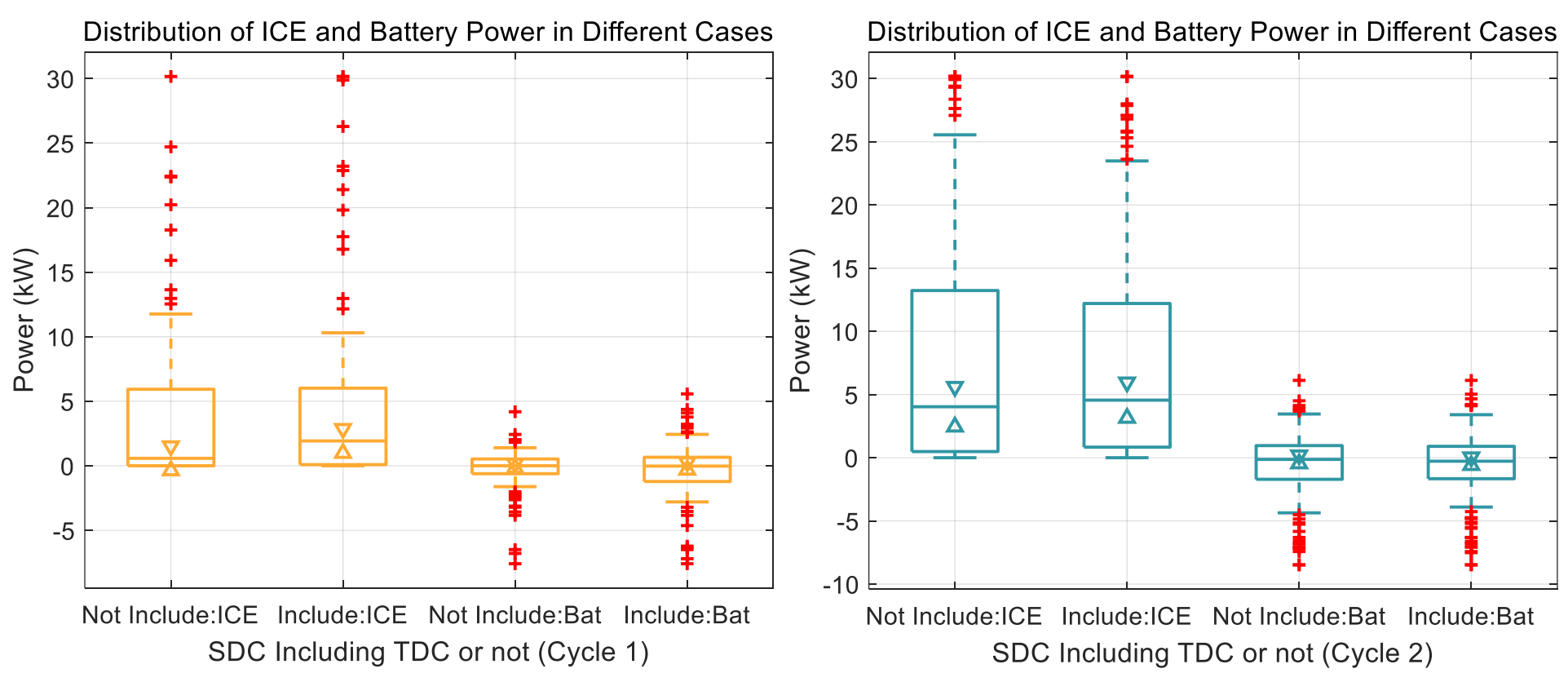

Fig. 10. Power distribution between ICE and battery in two control cases for two TDC.

Then, the power distribution between the two onboard energy sources (ICE and battery) is given in Fig. 10. The boxplot figure describes the minimum value, medium value, and maximum value of the ICE and battery power (Bat for short). For TDC 1, several characteristics can be observed. The ICE power is always greater than zero, and the battery power wanders in zero. It means that the battery could locate in charging or discharging condition. Furthermore, in the case of SDC including TDC, the medium power of ICE is more prominent than another case. It implies that the ICE is able to work at the high-efficiency area to achieve high fuel economy. For TDC 2, similar features could be discovered. Since the sum of the ICE and battery powers is the power demand, the medium value of battery power would be lower in the including case (means the SDC including the TDC information). This pattern of power distribution would affect the related reward, which will be exhibited in the following content.

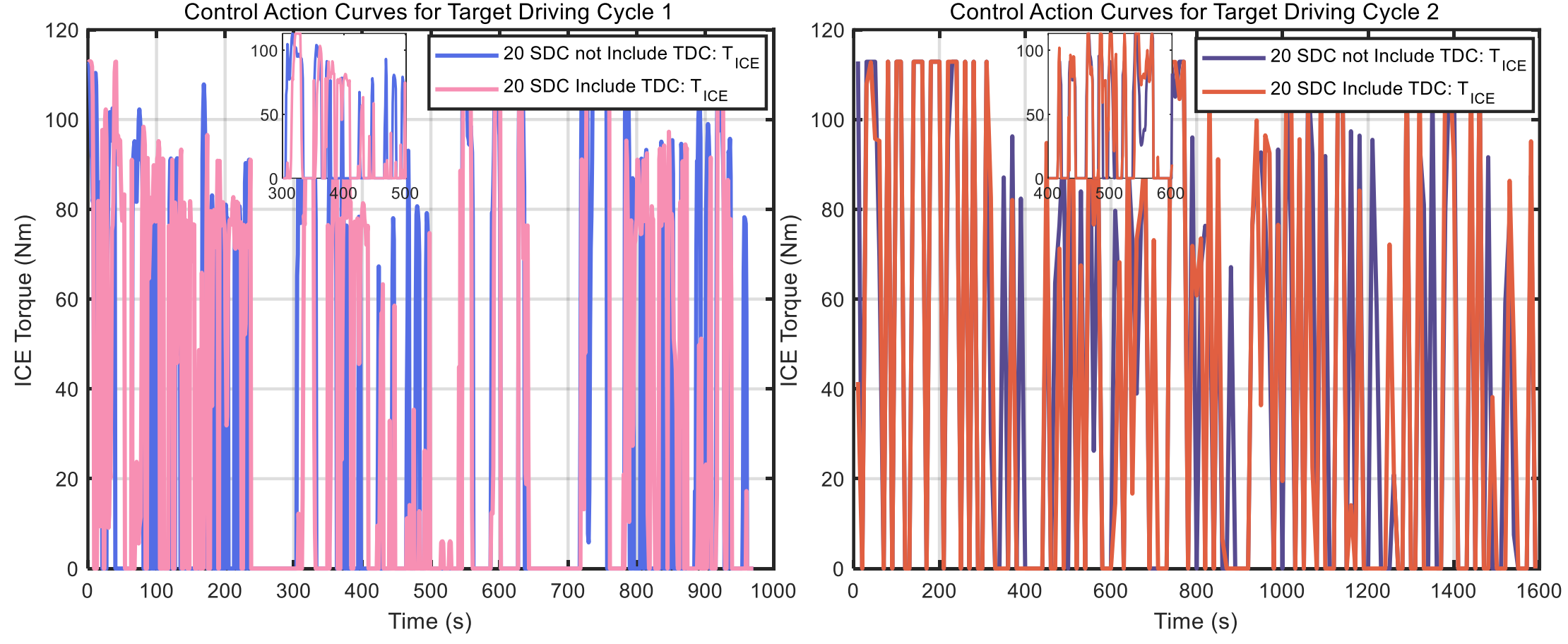

Fig. 11. Control strategies for two cases (SDC including TDC or not) for two TDC.

Furthermore, the power distribution is straightway influenced by the control action (ICE torque), and the relevant trajectories of ICE torque are depicted in Fig. 11. It is apparent that the working conditions of ICE are influenced by the driving cycles. If the vehicle speed is zero in Fig. 4, the ICE usually stops to provide the power, and the powertrain works in the electric mode (power is provided by the battery). In each TDC, the zoom-in graph shows the control policies of two control cases are different. It indicates the related EMSs for HEV are not the same. This feature reflects that the transformed information from the two source tasks is different. The corresponding reward would explain which transformed neural networks are better in the energy management problem.

Finally, the reward variations in two control cases of two TDC are sketched in Fig. 12. It can be discerned that the reward is lower when the vehicle speed is higher. At these time points, ICE needs to work to provide power, and the SOC decreases. From the cost function in (7), the reward would be lower at these time instants. For TDC 1, the total rewards in not including and including cases are -121.7403 and -118.0950, respectively. For TDC 2, the related rewards are -416.8146 and -405.101, respectively. It can be recognized that the including case could lead to better control performance. It states that the TDC information is able to improve the effectiveness of the transfer PPO-based EMS. In the energy management problem, future driving information (e.g., vehicle speed, acceleration, and road slope) is essential for the performance of EMS. The predictive techniques are considered to be incorporated in the proposed control framework to enhance the control performance in future work.

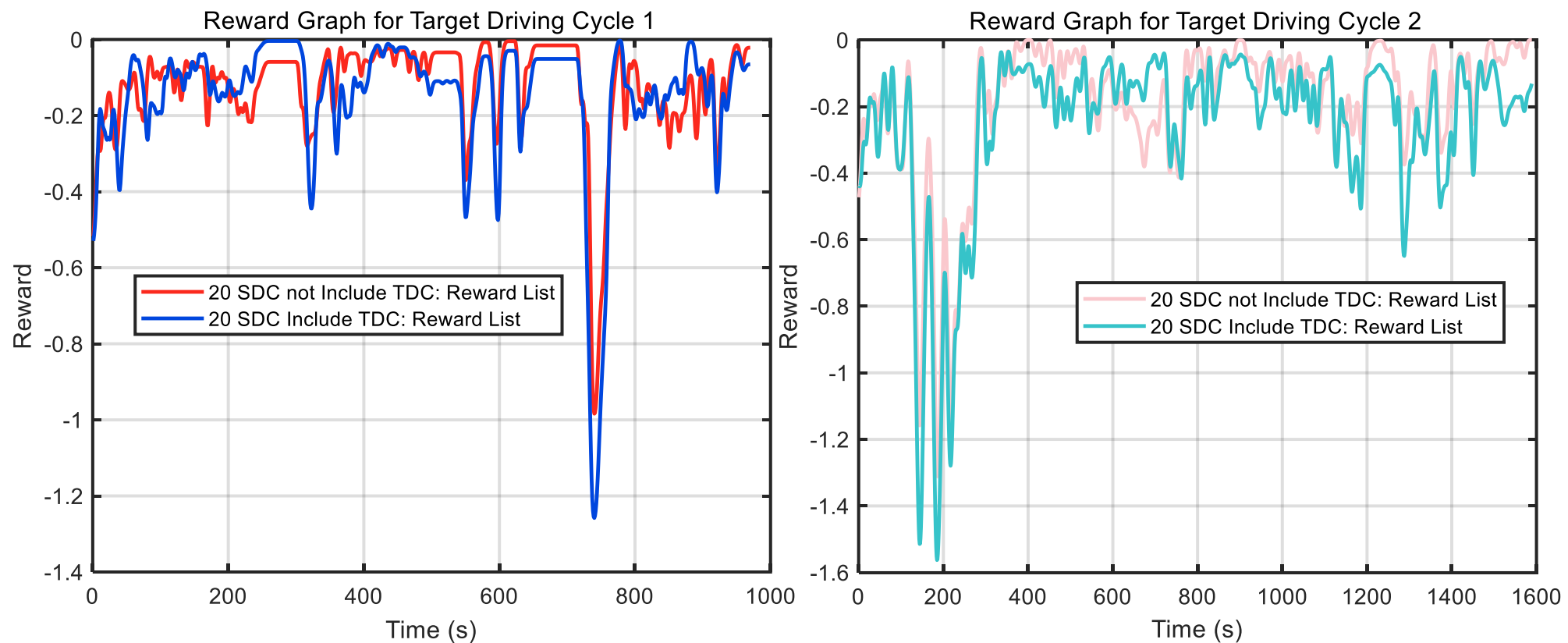

Fig. 12. Reward trajectories in the compared control cases with respect to two TDC.

### 4.3 Evaluation of TL Approach

In this part, the advantage of the presented TL method is explained. The PPO algorithm with and without the TL approach are compared in the energy management problem. In this work, the TL technique is leveraged to transform the trained neural networks. Based on the transformed parameters, the intelligent agent is capable of searching the optimal control actions as quickly as possible. Hence, the TL method is used to improve the learning efficiency of the DRL algorithm. In the transfer case, the number of SDC is still 20, and the TDC are two driving cycles shown in Fig. 4.

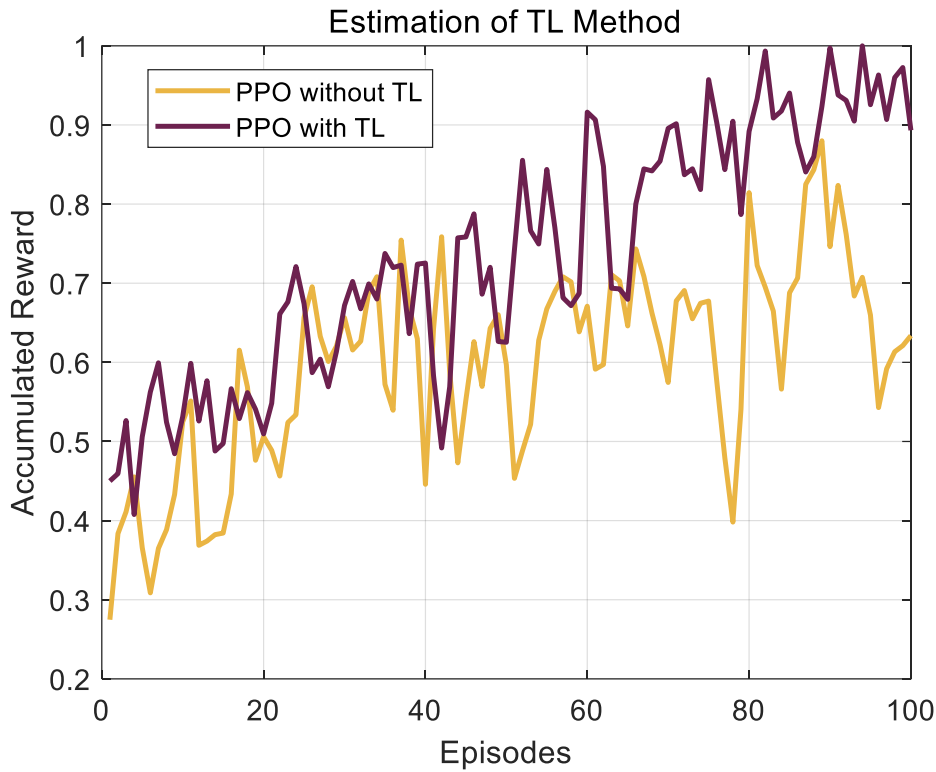

Fig. 13. Cumulative reward in two control cases: PPO with TL and PPO without TL.

The normalized cumulative rewards in the individual PPO algorithm and PPO with the TL algorithm are displayed in Fig. 13. The number of episodes is 100 in these two cases. It can be found that the starting point of reward in the PPO with the TL case is higher than that in PPO without TL. This is caused by the transformed neural network, which improves the efficacy of the chosen control actions. The accumulated reward curve in the PPO with the TL case is always higher than that in another case, which implies that the related control performance is better in the former case. It means the TL method could help the energy management controller to understand the powertrain dynamics better. As a result, the generated EMS is more appropriate for the HEV. By doing this, the final total reward in PPO with the TL case is higher than that of PPO without TL.

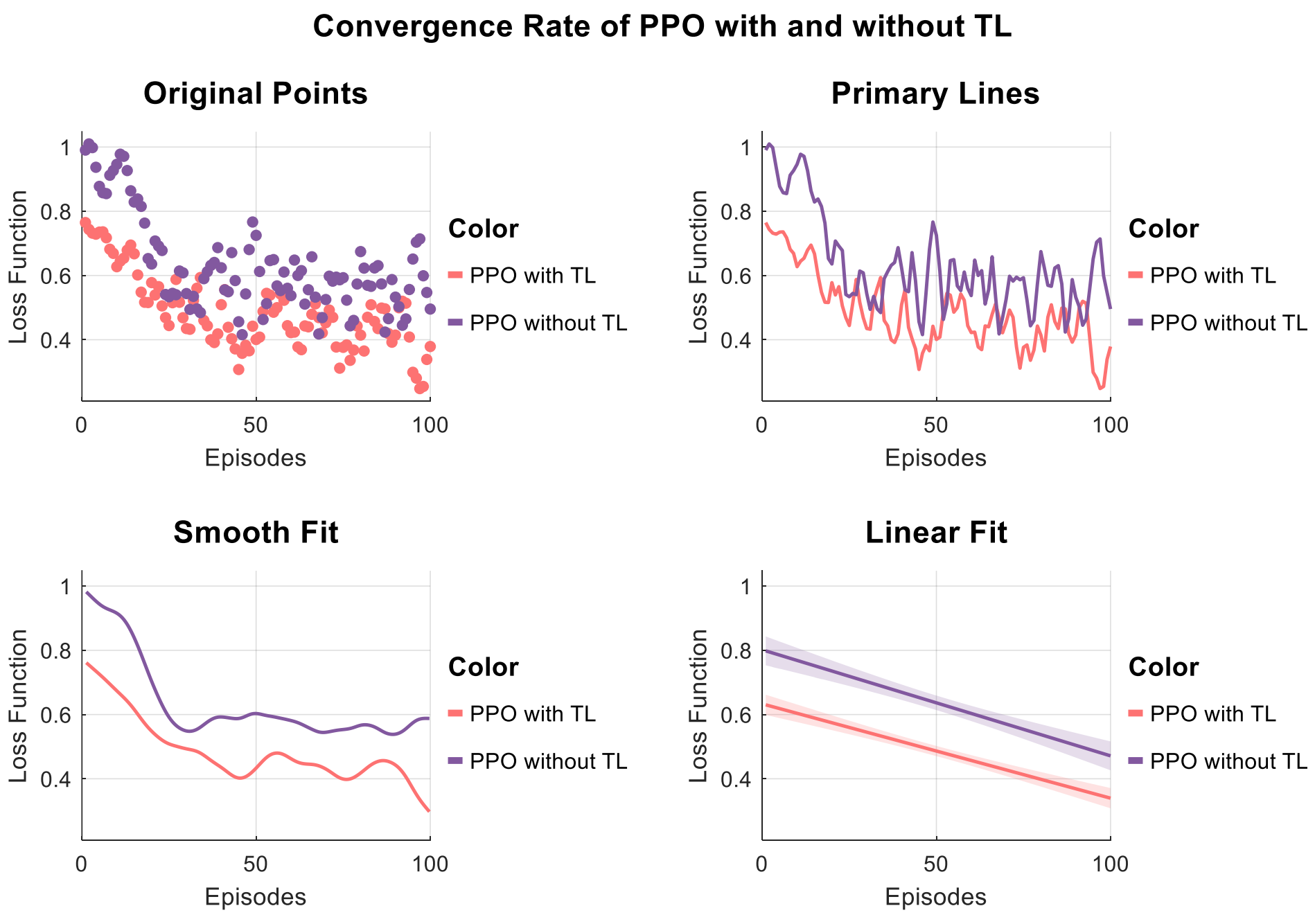

Fig. 14. Different types of the loss function to highlight the advantages of the TL method.

Another factor to represent the learning efficiency in DRL methods is the value of loss function. It indicates the convergence rate of the relevant algorithms. Four shapes of the loss function are given in Fig. 14, which are original points, primary lines, smooth approximation, and linear approximation. From this figure, the transformed parameters could reduce the initial value of the loss function. In each episode, the loss function in PPO with the TL case is lower than that in PPO without TL. This feature reflects a more mature Q-matrix could be achieved by PPO with TL. In the graphs of smooth approximation and linear approximation, the convergence rate in PPO with TL is obviously lower than that in PPO without TL. This appearance explains that the proposed transfer PPO-based EMS a faster convergence rate. Based on the testing results in Fig. 13 and 14, it can be concluded that the TL method is able to improve the learning efficiency and guarantee the control performance. To reduce the computational time, the proposed transfer PPO-based EMS has the potential to be applied in a real-time energy management controller.

## 5. Conclusion

A data-driven energy management policy is presented for HEV using the transfer DRL method in this work. The PPO algorithm is applied to address the energy management problem with continuous action space. The TL method is utilized to transform the learned neural networks from source tasks to target tasks. The relevant driving cycles are extracted from the real-world driving dataset. It implies that the derived EMS is able to be employed in real-time implementation. The simulation results discuss the impact of source driving cycles and target driving cycles on the EMS. The merits of the TL method are also emphasized.

Future works focus on two aspects to promote the proposed energy management framework for HEV. First, future driving information will be considered. The deep learning approaches are used to forecast the future vehicle speed and acceleration. This information would improve the fuel economy of the EMS. Second, the real-time application of the EMS is validated. The related hardware in loop (HIL) experiments will be conducted to estimate the online characteristics of the presented energy management policy for multiple hybrid powertrains.

**Acknowledgements**

The work was in part supported by National Nature Science Foundation of China (NSFC 51575064), State Key Laboratory of Mechanical System and Vibration (Grant No. MSV202016), and NNSF (Grant No. 11701027).